\documentstyle[aps,prl,multicol,epsf,psfig]{revtex}
\begin{document} 


 
\draft 
\title{Statistics of Coulomb Blockade Peak 
Spacings within the Hartree-Fock Approximation} 
%
%
\author{
Avraham Cohen$^{1}$, Klaus Richter$^{2}$, and 
Richard Berkovits$^{1}$
}
\address{
$^1$The Minerva Center for the Physics of Mesoscopics, 
Fractals and Neural Networks, Department of Physics, \\
Bar-Ilan University, 52900 Ramat-Gan, Israel \\ 
$^2$Max-Planck-Institut f\"ur Physik komplexer Systeme, 
N\"othnitzer Strasse 38, 01187 Dresden, Germany 
} 
\date{\today}
\maketitle 

\begin{abstract}
We study the effect of electronic interactions on the addition 
spectra and on the energy level distributions of two-dimensional 
quantum dots with weak disorder using the self-consistent 
Hartree-Fock approximation for spinless electrons.
We show that the distribution of the conductance peak 
spacings is Gaussian with large fluctuations that exceed, in
agreement with experiments, the mean level spacing of the 
non-interacting system. 
We analyze this distribution on the basis of Koopmans' theorem.
We show furthermore that the occupied and unoccupied Hartree-Fock
levels exhibit Wigner-Dyson statistics.
\end{abstract}




\begin{multicols}{2} 


\noindent 
{\em Introduction--- }
It has recently been established that studying the low temperature 
fluctuations of the conductance through quasi-isolated nanostructures
is an excellent tool for probing electronic interactions\cite{lee93}. 
In such small devices called artificial atoms\cite{mt90} or 
quantum dots it is possible to control and successively vary the total 
number of electrons from zero\cite{tarucha97} to a few 
hundreds\cite{sivan96,simmel97,patel98,simmel99}. 
When coupled weakly to leads, the conductance of the dots
is vanishingly small between pronounced peaks because of
Coulomb blockade\cite{kastner92meirav95}.
By tuning a gate voltage connected to the dot 
a conductance peak is observed whenever the ground state energy $E_G(n)$ of
the dot containing $n$ electrons becomes degenerate with the
energy  of the dot with $n\! +\! 1$ electrons. 
At almost zero drain to source voltage 
this situation is expressed as
%
%
%
%
$\mu_n \equiv E_G(n+1)-E_G(n) = \mu_{\rm Lead}$, 
where $\mu_{\rm Lead}$ and $\mu_n$ is the chemical potential of the leads
and of the dot containing $n$ electrons, respectively.
The spacing between adjacent conductance peaks 
is given by the difference 
%
\begin{eqnarray} 
s_n=\mu_n \!-\! \mu_{n-1}=E_G(n\!+\!1)-2E_G (n)+E_G(n\!-\!1) \; .
\label{sng} 
\end{eqnarray}  
 
Recently, a new generation of experiments on Coulomb blockade
in ballistic and diffusive quantum dots has shown that the
conductance peaks fluctuate 
both, with respect to their heights\cite{j92folk96chang96} 
and spacings\cite{sivan96,simmel97,patel98,simmel99}. 
Most of the features of the peak {\em height} statistics are
well understood within random matrix theory (RMT)\cite{jalabert92}.
However, all measured\cite{sivan96,simmel97,patel98,simmel99} 
peak {\em spacing} distributions resemble a Gaussian form, while
the constant interaction model\cite{kastner92meirav95}, 
which properly accounts for the average peak spacing, together with RMT 
predicts a Wigner-Dyson statistics. Moreover, the observed widths
of the distributions vary between the experiments from a width 
comparable to the mean single-particle spacing $\Delta$\cite{patel98} 
to a width considerably larger\cite{sivan96,simmel99},
whereas the widths of the GOE and GUE distributions are about $\Delta/2$.
These experiments suggest that the widths scale rather with the
charging energy $E_c$ than with $\Delta$.

Different theoretical studies addressed this issue:
RPA diagrammatic perturbation theory yields 
corrections to the constant interaction predictions of order $1/g$ 
\cite{berko97} or $1/\sqrt{g}$ \cite{blanter97}, where $g$ is the 
dimensionless conductance. However, these approaches,
which are valid at small $r_s$ (high densities), 
cannot easily explain the large widths ($\sim 8 \Delta$)
of the latest experiments\cite{simmel99}.
A refined RMT approach\cite{vallejos98}, which 
accounts for shape deformations of the dot while adding electrons, 
explains the Gaussian profiles, while the widths remain
comparable with $\Delta$.
 
Numerical exact diagonalization studies of an interacting tight-binding
model of $\sim\! 10$ electrons 
show that 
for $r_s\!=\!1$ the distribution is indeed Gaussian \cite{sivan96}.
This is also the
situation when spin is included\cite{stopa,berko98}. Nevertheless, questions
were raised regarding the relevance of these calculations to the experimental
system due to the small number of electrons considered, the use of a
tight-binding model for which it is difficult to relate the parameters to
the experimental systems, and due to the small values of $g$ (of order one) 
accessible in this approach.

In this paper we study interacting spinless electrons on continuous 
two-dimensional disordered cylinders at zero temperature. The
interactions are considered within the self-consistent Hartree-Fock (SCHF) 
approximation, which enables us to treat up to $n\!=\!50$ electrons. 
By using a continuous model, the experimental parameters can directly 
be related to the calculations, and we obtain higher values of $g$ 
($g \sim 10$) close to the experimental values. 
We show that within the SCHF approximation
the peak spacing statistics is Gaussian in
the presence or absence of magnetic field. The  variance
of the fluctuations is much larger than predicted in the
independent-particle picture. This is in agreement with the exact 
diagonalisation results, with a density functional approach\cite{stopa},
and with other recent HF studies of 
disordered tight-binding models\cite{levitgefen} and 
clean quantum dots\cite{ahn99}. Using Koopmans' theorem\cite{koopmans35}
we show that the calculated widths and Gaussian profiles can,
at least partly, be related to specific Hartree interaction
integrals (for not too strong interactions). 
We further show and explain that, on the contrary,
the spacings among the occupied or the unoccupied HF levels 
follow GOE and GUE level statistics.

{\em Numerical procedure---}
We begin with the HF equations for 
spinless electrons in real space,
\begin{eqnarray} 
&  & 
\left[ H_0  +
V_{\rm dis}(\vec{r}) + {{e^2}\over{4\pi \epsilon}} 
\int_{\Omega '}  {   \sum_{j=1}^{n} | \psi_{j}(\vec{r'}) |^2
\over |\vec{r}-\vec{r'}|   } 
d{\Omega '} \right] \psi_{k}(\vec{r}) -  \nonumber \\
&  & -
{{e^2}\over{4 \pi \epsilon}}  
\int_{\Omega '}
{ \sum_{j=1}^{n}\psi_{j}^{*}(\vec{r'})\psi_{k}(\vec{r'}) 
\psi_{j}(\vec{r})  
\over |\vec{r}-\vec{r}'|  }   d\Omega '
 =  
\varepsilon_k \psi_{k}(\vec{r})  \; .
\label{hfintegro}
\end{eqnarray} 
%
%
%
%
Here $\epsilon$ is the dielectric constant, 
$H_0= - (\hbar^2/ 2m_e^*) \partial^2/  \partial \vec{r}^2$ and
$V_{\rm dis}(\vec{r}) \!=\! 
\sum_{q=1}^{N_s} \lambda_q\delta(\vec{r}-\vec{r}_q)$ 
the potential from $N_s$ $\delta$-scatterers at 
random locations $\vec{r}_q$ homogeneously 
distributed on the cylinder surface $\Omega$. 
The strengths $\lambda_q$ are chosen from a 
box distribution.
%
 
We use $0\! \le\! \theta\! \le\! 2\pi$ as polar and  
$0 \le y \le \pi$ as vertical coordinates
and boundary conditions,
$\psi_j(\theta,0)\!=\!\psi_j(\theta,\pi)\!=\!0$,  
$\psi_j(\theta+2\pi, y)\!=\!e^{i2\pi\phi / \phi_0} \psi_j(\theta, y)$,
to account for a flux $\phi$ piercing the cylinder. 
$\phi_0\!\equiv\!hc/e$ is the flux quantum.    
The set of Eqs.\ (\ref{hfintegro}) is solved 
self-consistently by diagonalizing the Hamiltonian 
matrix in the basis of $H_0$. 
To this end  we expand 
$\Psi_j(\vec{r})\equiv e^{-i\theta \phi/ \phi_0}
\psi_j(\vec{r}) = \sum_{l} a_{l}^{(j)} u_{l}(\vec{r})$   
in the periodic basis functions 
$u_{l}(\vec{r})=(1/ \pi)e^{in_\theta \theta}\sin(n_y y)$ of $H_0$,
where $n_\theta=0, \pm 1, \pm 2, ...$ and $n_y=1,2,3...$ are good quantum 
numbers which define a level $l=l(n_\theta,n_y)$ of $H_0$.
We emphasize that the enumeration of the levels $l(n_\theta,n_y)$ 
does depend on flux. 
Using energy units $\hbar^2/ (m_e^*R^2)$, 
where $R$ is the cylinder radius, the matrix elements of the
Hamiltonian (\ref{hfintegro}) read
\begin{eqnarray}
H(l_1,l_2) = \varepsilon_{l_1}^0 \delta_{l_1,l_2}\! + \!V_{\rm dis} (l_1,l_2) 
+ 
U\! \sum_{l_1',l_2'} \! A_{l_1',l_2'} V_{l_1',l_2'}(l_1,l_2)\; .\nonumber     
\label{hfmatrix}
\end{eqnarray}  
Here, $l,l_i,l_i'$ denote eigenstates of $H_0$ and
$\varepsilon_l^0\!=\! 1/2 \left[( n_\theta + 
\phi/\phi_0 )^2 \!+\! (n_y/ 2)^2\right]$  
is the scaled energy of an electron on a clean cylinder of 
aspect ratio unity. 
$V_{\rm dis} (l_1,l_2)$ are the disorder matrix elements.
$U \equiv R/2r_B$ 
determines the interaction strength in dimensionless units with
$r_B\equiv 4 \pi \epsilon \hbar^2/ (m_e^* e^2)$ 
the Bohr radius in the sample material. 
$A_{l_1',l_2'} = \sum_{j=1}^{n} a_{l_1'}^{*(j)}  a_{l_2'}^{(j)} $ 
and 
$V_{l_1',l_2'}(l_1,l_2) = (l_1,l_1'|v|l_2,l_2') - (l_1,l_1'|v|l_2',l_2)$
with $v \equiv 1/|\vec{r} - \vec{r'}| = \left[\sin^2(\theta - \theta')/
2) +(y-y')^2\right]^{-1/2}$ in real space.   
While the matrix $A$ changes along the iterations until self-consistency
is reached,  $V_{\rm dis}$ is calculated only once.  
We used $450 \times 450$ matrices and ensured
that the results are practically unaffected by this truncation.

Generally, interaction effects are governed 
by the charge density $n_s$ and by  $U$.   
We used $U=0.2\pi^2=1.97$. On the one hand, this
choice still allows convergence of the SCHF, on the other hand 
this is in the experimental parameter regime
\cite{sivan96,simmel97,patel98}: In GaAs 
$r_B =10^{-2} \mu {\rm m}$ 
($m_e^*=0.067 m_e$, $\epsilon=13 \epsilon_0$). This gives
for a typical dot area of $(2 \pi R)^2 \sim 0.07 - 0.3 \mu {\rm m^2}$  
the estimate $U \sim 2 - 4$ which is similar to the numerical value used.
Accordingly, 
typical experimental values of $r_s \sim 1$ for GaAs 
\cite{sivan96,simmel97,patel98} and $r_s \sim 2$ for Si \cite{simmel99}
are also close to our
numerical value $r_s=2 U \sqrt{\pi/n}$ that 
varies between 1.0 and 1.6 for $20\le n \le 50$ electrons.  
The disorder strength used
corresponds to $g \sim 10$ describing the experimental 
situation. $g$ was calculated from the
inverse participation ratio\cite{prigodin}.

{\em Results and analysis---}
For a system containing $n$ electrons the SCHF ground state energy
is given by\cite{heinonen}   
\begin{eqnarray}
E_G(n) = \sum_{k=1}^{n} \varepsilon_k - {1 \over 2} 
\sum_{j,k=1}^{n} \left[ (j,k|v|j,k) -  (j,k|v|k,j) \right]  \; ,
\label{Eg}
\end{eqnarray}  
where $j$ and $k$ enumerate SCHF levels and
we redefined $v \equiv U / |\vec{r}  -\vec{r'}|$.
Eq.~(\ref{Eg}) was used to calculate the peak spacings 
$s_n$ in Eq.~(\ref{sng}). The peak spacing distributions were generated
from an ensemble of 600 quantum dots (with different disorder realizations)
and sequences of $s_n$ with $20 \! <\! n \!< 50$. The results are
depicted as the histograms 1 in Fig.~1 for $\phi = 0$  and Fig.~2 for 
finite $\phi$. The curves are centered
around the mean peak spacing $\langle s_n \rangle \simeq 8.3 \Delta $.
Clearly each distribution fits a Gaussian, curves 3, defined by the
average and variance of the data. 

In the following we analyze the numerical results.
Assuming that the single-particle states are unchanged by adding (subtracting) an 
electron to (from) the many-electron system we can
approximate the difference in energy of the ground states
of the system with $n+1$ and $n$ electrons 
according to Koopmans\cite{koopmans35} by
\begin{equation}
\mu_n \equiv E_G(n+1) - E_G (n) \simeq \varepsilon_{n+1} (n) \; .
\label{koopmans}
\label{mu}
\end{equation}
Here, $\varepsilon_{n+1} (n)$ is the $n\!+\!1$st energy
eigenvalue in the system of $n\!\gg\! 1$ electrons. 
In view of Eq.~(\ref{sng}), we then have approximately 
\begin{eqnarray}  
s_{n} \simeq  \varepsilon_{n+1} (n) - \varepsilon_{n} (n-1) \; .
\label{snk} 
\end{eqnarray} 
The numerical distributions of the $s_n$ from Eq.~(\ref{snk}) are
depicted as curves 2 in Figs.~1 and 2. They show considerable
agreement with the corresponding distributions (curves 1) of the SCHF
ground state energies, Eq.~(\ref{sng}), and their Gaussian fits.
This shows that, for the interaction strengths considered, a
peak spacing analysis based on Koopmans' relation is justified.
As visualized in the right inset of Fig.~1, the peak
spacing is related to energy differences from systems with 
different number of electrons\cite{nogap}. 
We emphasize that $\varepsilon_{n} (n\!-\!1) \ne \varepsilon_{n}(n)$. 
Their difference is typically large (50$\%$ and more) 
on the scale of $s_n$ in Eq.~(\ref{snk}). Hence
$s_n \ne \varepsilon_{n+1} (n) - \varepsilon_{n} (n)$.

We also checked directly the validity of Koopmans' relation (\ref{mu})
by comparing the two numerical values for $\mu_n$ and $\epsilon_{n+1}(n)$.
The typical relative error, the difference 
$ (\varepsilon_{n+1} (n) - \mu_{n})/\mu_n$, is always positive
and  $\sim 0.4\%$, indicating
the validity of Eq.~(\ref{mu}).
%

\begin{figure}
[t]
\psfig{figure=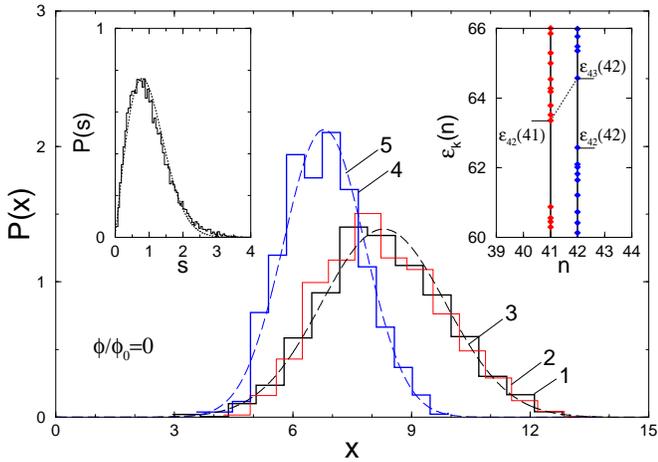,angle=-90,height=6.4cm} 
\caption{   
Distributions of normalized Coulomb blockade peak spacings 
$x\!=\!s_n / \Delta$ for weakly disordered cylindrical quantum
dots and  flux $\phi\!=\!0$. The histograms 1 and 2 are
calculated from the self-consistent Hartree-Fock ground state 
energies (Eq.~(\ref{sng})) and by using Koopmans' relation (\ref{snk}). 
Curve 3 is a Gaussian defined by the average (the charging
energy) and variance of the data of curve 1. 
Curve 4 is the distribution of the exceptional direct 
integral (\ref{dirint}) yielding a considerable
contribution to $s_n$ in Eq.~(\ref{snekpmn}). 
Curve 5 is a Gaussian fit.
{\bf Left inset:} 
The nearest neighbour spacing distribution of SCHF single-particle 
energies for the occupied and unoccupied states 
is compared with the  GOE statistics
$p(s)=(\pi/2)s \exp(-\pi s^2/4)$ (dotted line). 
{\bf Right inset:} 
A typical realization of SCHF eigenvalues. The dotted
line connects eigenvalues relevant for
calculating e.g., $s_{n=42}$ using Koopmans' relation (\ref{snk}). 
}
\label{fig1}
\end{figure} 
%
\begin{figure}
[t]
\vspace{-0.3cm} 
\psfig{figure=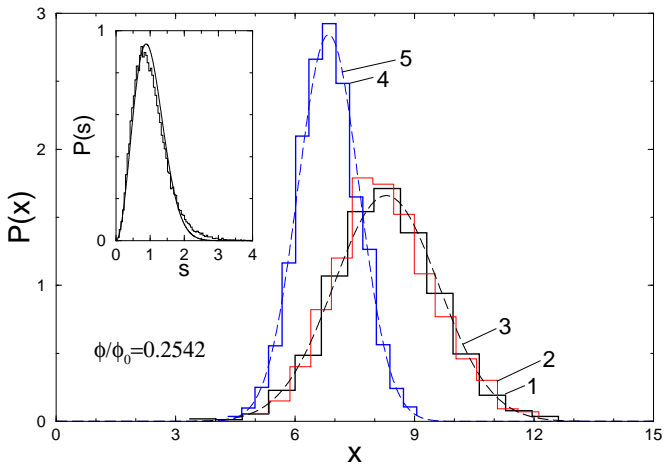,angle=-90,height=6.4cm} 
\caption{   
SCHF peak spacing distributions as in Fig.~1 but for 
$\phi /\phi_0\simeq$ 0.25 which reduces the fluctuations.  
{\bf Inset:}  
The nearest neighbour  distribution, 
as in the left inset of Fig.~1, 
of SCHF single-particle energies  
is compared with the GUE distribution 
$p(s)=(32 / \pi^2) s^2 \exp(-4s^2/\pi)$ (dotted line). 
}
\label{fig2}
\end{figure} 

To understand the origin of the fluctuations of $s_n$ 
we recall that the HF energies $\varepsilon_k$ depend on $n$ 
and obey\cite{heinonen}   
\begin{eqnarray}
\varepsilon_k(n) = 
(k| H_0\!+\!V_{dis} |k) +\! 
\sum_{j=1}^n (j,k|v|j,k) \!-\! (j,k|v|k,j).
\label{ek}
\end{eqnarray}  
Using this relation in Eq.~(\ref{snk}) we obtain
\begin{eqnarray} 
s_{n}   
&\sim&   
\left[ (n+1|H_0+V_{dis}|n+1) - (\tilde{n}|H_0+V_{dis}|\tilde{n}) \right] 
\nonumber \\ 
&+&  
(n,n+1|v|n,n+1) - (n,n+1|v|n+1,n) \nonumber \\ 
&+&  
\sum_{j=1}^{n-1} \left[  (j,n+1|v|j,n+1) - 
(\tilde{j},\tilde{n}|v|\tilde{j},\tilde{n}) \right] 
\nonumber \\ 
&-&     
\sum_{j=1}^{n-1}\left[ (j,n+1|v|n+1,j) - 
(\tilde{j},\tilde{n}|v|\tilde{n},\tilde{j}) 
\right] . 
\label{snekpmn} 
\end{eqnarray}   
The tilde sign denotes states calculated in the system with
$n\!-\!1$ electrons. 
When $v\rightarrow 0$ one recovers the  Wigner-Dyson
distribution for $s_{n}$ which is modified upon increasing the
interaction.
The two sums in Eq.~(\ref{snekpmn}) contain differences between
direct terms and between exchange terms which partly cancel each other.
This is not the case for the direct and exchange integral in
the second line which therefore represents an exceptional
contribution to $s_n$.
The direct term
dominates over the exchange term
because the integrand of the former 
%
is always positive.  Hence we expect that the exceptional direct integral,
$(n,n+1|v|n,n+1)$, describing the interaction between the $n$-th and 
$n\!+\!1$st electron and given by 
\begin{eqnarray} 
U \int_{\Omega}\int_{\Omega'} 
{|\Psi_{n}(\vec{r})|^2
|\Psi_{n+1}(\vec{r}')|^2   
\over |\vec{r}  -\vec{r'}|} d\Omega d\Omega '     \; ,
\label{dirint} 
\end{eqnarray} 
significantly contributes to $s_n$. 

It has been established\cite{berry77,prigodin93prigodin95} 
that for moderate disorder the eigenfunctions 
fluctuate in space with small correlations.
They even survive for different eigenfunctions 
but are weaker\cite{blanter97mirlin97}.   
This should hold also in our SCHF calculations because the 
screening of the disorder potential is not perfect and 
disorder effects will persist.
Because of the long-ranged Coulomb term,
the central limit theorem should hold fairly good for 
the direct (Eq.~(\ref{dirint})) and exchange integrals 
(despite the small wave functions correlations) and
one thus expects to find a Gaussian distribution.
For the interaction strengths used ($U\simeq 2$) the 
direct term (\ref{dirint}) can dominate the $s_n$ 
which implies a Gaussian distribution for the peak spacings, too.

In Figs.~1 and 2 the numerical distribution of the direct 
term (\ref{dirint}) is shown as curve 4 and is compared 
with a Gaussian fit, curve 5. This distribution is 
expected to be centered to the left of the full numerical
spacing distribution, because we have also to account for
a shift on the scale of the mean HF single-particle spacing $\Delta_{HF}$
according to the first line in Eq.~(\ref{snekpmn}).
%
Note that below the highest occupied and above
the lowest unoccupied HF level, the spacing is 
$\Delta_{HF}=0.314 > \Delta=0.183$.   
The shift to the left in Fig.~1 and Fig.~2    
is $\sim 0.280 \sim 0.88 \Delta_{HF}$ and 
$\sim 0.270 \sim 0.84 \Delta_{HF}$, respectively.

Due to the appreciable interaction strength 
the discussion based on Eq.(\ref{snekpmn}) 
is practically independent of flux. Both $s_n$ statistics 
are close to Gaussians but with flux-dependent widths. 

{\em Comparison with experiment--- }   
The Gaussian form of the distributions is in good agreement 
with all experimental results\cite{sivan96,simmel97,patel98,simmel99}. 
The average peak spacing of the interacting systems is 
$\langle s_n\rangle /\Delta\!=\!1.52/0.183\!=\! 8.27 \gg 1$ 
(curve 1 in Figs. 1 and 2),
similar to the experimental situation in GaAs \cite{sivan96,simmel97,patel98}.
$\langle s_n\rangle $ is also in agreement with the estimated (scaled) classical
capacitance of the cylinder, 
$[e^2/4\pi \epsilon (\pi R)]/[\hbar^2/m_e^*R^2] = 0.4 \pi = 1.26$ 

The width (RMS) of curve 1 in Fig.~1 (Fig.~2) equals 
 $ 1.6 \Delta = 0.19 \langle s_n\rangle$ ($ 1.3 \Delta = 0.16 
\langle s_n\rangle $).
This is considerably larger than the widths 
$0.0966= 0.52 \Delta$ 
($0.0743= 0.41 \Delta$) for the non-interacting case which
are in good agreement with GOE (GUE) predictions\cite{berkovits}.

The enhancement of the widths is compatible with the first 
GaAs experiments\cite{sivan96,simmel97} and in line
with the Si experiment\cite{simmel99} which shows a larger 
width for higher values of $r_s$.
The width is large
compared to the experiment \cite{patel98} where it
was close to its non-interacting value. 
Since all experimental settings are quite similar, 
it is not clear what can lead to the different widths
observed in\cite{patel98}. 

{\em SCHF single-particle statistics--- }
We finally consider the statistics of the
HF level spacings $s_k$ with $k\! \ne\! n$ to emphasize that
they are {\em not} related to the Gaussian spacing distribution.
By using Eq.~(\ref{ek}) we have 
\begin{eqnarray} 
s_{k} &=& \varepsilon_{k+1}(n)-\varepsilon_k(n) \nonumber \\    
&=&  (k+1|H_0+V_{dis}|k+1) - (k|H_0+V_{dis}|k) \nonumber \\ 
&+&  
\sum_{j=1}^{n}\left[ (j,k+1|v|j,k+1) - (j,k|v|j,k)\right]\nonumber \\ 
&-&     
\sum_{j=1}^{n}\left[ (j,k+1|v|k+1,j) - (j,k|v|k,j) \right] .  
\label{sk} 
\end{eqnarray}   
In contrast to Eq.~(\ref{snekpmn}) the exceptional and 
dominant direct integral does not appear in Eq.~(\ref{sk}) 
leading to $s_k \ll s_{n}$. 
It is easy to see that for $k>n$ all terms contribute and
for $k<n$  all terms except those with $j=k$ and $j=k+1$ that 
cancel each other. 
Hence there is no essential difference in the
$s_k$ distributions in the two $k$ regions.
Because of the tendency of 
the Coulomb terms to cancel each other, the terms in the
second line in Eq.~(\ref{sk}) are expected to dominate the average. 
The left insets of Figs.~1 and 2 demonstrate that the numerical 
distributions of the normalized $s_k$ follow indeed GOE and GUE
statistics (dotted lines), respectively, showing
the effect of time-reversal symmetry breaking.
This RMT type of statistics is in line with other HF results 
that was found in a different type of experiment\cite{sivan94}.
 
{\em Conclusion--- }
We have studied Coulomb blockade peak spacing fluctuations
for interacting spinless electrons within the self-consistent Hartree-Fock
approximation. The main features of the conductance peak spacing 
fluctuations are similar to the experimental ones. 
The Gaussian distribution of the addition spectrum is 
interpreted as the result of the dominance of direct 
interaction terms governed by spatial fluctuations 
of the eigenfunctions. 
On the other hand the Hartree-Fock single-particle levels
were shown to follow Wigner-Dyson statistics.  

%

%

One of us, AC, would like to thank D.\ Bar-Moshe, A.\ Heinrich, 
S.\ Levit, D.\ Orgad, B.\ Shapiro and S.\ Shatz for valuable 
discussions. It would not have been 
possible to complete this work without crucial consultation
in super computations by N.\ Fraenkel, G.\ Koren and J.\ Tal 
from the IUCC at Tel-Aviv University and S. Shatz 
from BIU.



\end{multicols}


\begin{references}
\bibitem{berkovits}
This is very similar to the exact diagonalization 
results\cite{sivan96,berko98} for which the fluctuations were 
substantially enhanced compared to the non-interacting case, 
and for $r_s \sim 1$ were of order of $15\%$ of the average
peak spacing.
\end{references}
\end{document}